\PassOptionsToPackage{table}{xcolor}
\documentclass[]{svjour3} %
\usepackage[utf8]{inputenc}
\usepackage{todonotes}

\usepackage{natbib}
\usepackage{booktabs}
\usepackage{graphicx}
\usepackage[hidelinks]{hyperref}
\usepackage{appendix}
\newcommand{\bi}{\begin{itemize}}
\newcommand{\ei}{\end{itemize}}
\newboolean{showcomments}
\setboolean{showcomments}{false} %
\ifthenelse{\boolean{showcomments}}
    {
             \newcommand{\response}[2]{\textcolor{blue}{{\it \\ \noindent \textbf{Response #1}: #2 \\ }}}
        \newcommand{\martin}[1]{\textcolor{cyan}{{\it [Martin says: #1]}}}
         \newcommand{\neil}[1]{\textcolor{red}{{\it [Neil says: #1]}}}

    }
    {
        \newcommand{\martin}[1]{}
        \newcommand{\neil}[1]{}
        \newcommand{\response}[1]{}

    }
\newcommand{\replPackage}[0]{\url{https://zenodo.org/record/7539493}}
\journalname{Empirical Software Engineering}

\newcommand{\tool}{JetUML}%
\newcommand{\A}{McGill University}%
\newcommand{\B}{University of Victoria}%
\usepackage{multicol}
\usepackage{subcaption}
\usepackage{siunitx}
\usepackage[many]{tcolorbox}
\newtcolorbox{myboxi}[1][]{
  breakable,
  title=#1,
  colback=white,
  colbacktitle=white,
  coltitle=black,
  fonttitle=\bfseries,
  bottomrule=0pt,
  toprule=0pt,
  leftrule=3pt,
  rightrule=3pt,
  titlerule=0pt,
  arc=0pt,
  outer arc=0pt,
  colframe=black,
}

\usepackage{multirow}

\begin{document}

\title{A Study of Documentation for Software Architecture}
\author{Neil A. Ernst \and Martin P. Robillard}
\date{January 2023}
\institute{University of Victoria, Canada \and McGill University, Canada}

\maketitle

\begin{abstract}
Documentation is an important mechanism for disseminating software architecture knowledge. Software project teams can employ vastly different formats for documenting software architecture, from unstructured narratives to standardized documents. We explored to what extent this documentation format may matter to newcomers joining a software project and attempting to understand its architecture. We conducted a controlled questionnaire-based study wherein we asked 65 participants to answer software architecture understanding questions using one of two randomly-assigned documentation formats: narrative essays, and structured documents. We analyzed the factors associated with answer quality using a Bayesian ordered categorical regression and observed no significant association between the format of architecture documentation and performance on architecture understanding tasks. Instead, prior exposure to the source code of the system was the dominant factor associated with answer quality. We also observed that answers to questions that require applying and creating activities were statistically significantly associated with the use of the system's source code to answer the question, whereas the document format or level of familiarity with the system were not. Subjective sentiment about the documentation format was comparable: Although more participants agreed that the structured document was easier to navigate and use for writing code, this relation was not statistically significant. We conclude that, in the limited experimental context studied, our results contradict the hypothesis that the format of architectural documentation matters. We surface two more important factors related to effective use of software architecture documentation: prior familiarity with the source code, and the type of architectural information sought.
\end{abstract}

\section{Introduction}

The various forms of developer turnover and the continual evolution of software projects often create situations where \emph{newcomers} to a project must learn about the landscape of the project~\citep{DOB2010a, LRS2017a}. Such situations include the hiring of new developers, freelancers and consultants, internal transfers, and volunteers attempting to join open-source projects. Research into these \emph{onboarding} scenarios reports how developers face numerous challenges when joining a new project~\citep{SGG2015a}. One particular challenge is that of orientation: how to become familiar with the various features of the project, and in particular the \textit{architecture} of the software~\citep{DOB2010a}.

An important role for a software architecture is to facilitate communication between stakeholders~\citep{RW2012}. Consequently, a significant portion of the literature on software architecture focuses on human aspects~\citep{TRP2017a}, and in particular on documenting and sharing knowledge about large and complex software systems~\citep{RW2012,Brown:2012aa,Clements:2010aa,BRF2007a}. 

A software's architecture concerns its high-level abstractions, namely, the ``fundamental concepts or properties of the system in its environment, embodied in its elements, relationships, and the principles of its design and evolution''~\citep{RW2012}. To be useful, software architecture knowledge must thus be explicitly captured and represented as \textit{software architecture documentation}~\citep{Clements:2010aa}. Different frameworks have been proposed to organize architecture documentation, typically based on the concept of \textit{architectural views}~\citep{KRU1995a}, which are different projections of knowledge about a system (see Section~\ref{ss:SoftwareArchitectureDocumentation}). Despite the existence of well-supported frameworks for documenting software architecture, there also exist many examples of documentation of the software architecture for a project that does not follow any pre-determined format, and simply presents and explains architectural decisions as a narrative complemented with informal diagrams~\citep{Brown:2012aa,SG2009a}. To date, little empirical research has focused on assessing the impact of the way we represent software architecture knowledge, in particular for the purpose of initial orientation into a project.

We conducted a study to explore \textit{to what extent the format of architectural documentation mattered when joining a new software project}. In the study, we asked upper-year students in software architecture courses 
at two different universities and with different degrees of knowledge of a target system to read software architecture descriptions. The architectural descriptions captured the same information but were expressed in two different formats: one in the highly-structured \textit{Views and Beyond style}~\citep{Clements:2010aa}, and one in the free-form \emph{essay style} described by~\cite{Robillard2016}. We then asked the participants to answer questions that assessed their understanding of the architecture of the system, and further surveyed them about their experience using the documentation. We rated the participants' answers on a four-point ordinal scale and, using a regression model, we modeled the relation between this measure of performance, self-assessed knowledge of the code, and documentation format.

Our results show that, in the context of a first approach to the architecture of a comparatively small software system with about 1000 words of architectural documentation, the documentation format had little relationship to the measured level of understanding of the architecture. 
Instead, basic familiarity with the system's source code had a much more significant association with measured performance in answering questions about the architecture. 
In terms of sentiment, participants using the structured format were sightly more positive, in particular about the ease if navigating the document.  

This article is organized as follows. We begin with a survey of related work in architecture knowledge management, architecture documentation, and research on onboarding newcomers in software development projects (Section~\ref{s:background}). We then describe how we designed and conducted the study (Section~\ref{s:design}), describe the analysis methods and present our findings (Section~\ref{sec:findings}), and conclude (Section~\ref{s:conclusion}).

\section{Background and Related Work}
\label{s:background}

Our study is informed by background on \textit{architecture knowledge management} and \textit{software architecture documentation}, while addressing specifically the need to \emph{onboard developers into project teams}.

\subsection{Architecture Knowledge Management}

\cite{CJT2016a} provide a broad overview of the field of architecture knowledge management, with an extensive discussion of the needs for capturing software architecture knowledge. These needs had been previously organized into four broad functions for representations of architecture knowledge: \textit{Sharing} information, ensuring \textit{compliance} with a design, assisting the \textit{discovery} of new information, and supporting \textit{traceability}~\citep{BRF2007a}. Clerc et al.'s survey emphasizes the importance of architecture knowledge for creation and communication tasks~\citep{CLV2007a}. 

Software architecture knowledge can take many forms. For example, one can distinguish between \textit{general knowledge}, \textit{context knowledge}, \textit{reasoning knowledge}, and \textit{design knowledge}. \cite{GB2014a} report on the extent to which these different types of knowledge are evidenced among 29 industrial projects, finding that all knowledge types are found in a majority of the studied projects.

In this work, we construe software architecture knowledge in terms of what can be done with that knowledge. Others have conceptualized architecture knowledge differently. For example viewing documentation through a knowledge management  lens~\citep{Ding2014KnowledgeBasedApproaches}, with a distinction between producer and consumer of the documentation. The notion of \textit{understanding} is also much broader than looking at documents; it includes understanding people and processes as well~\citep{Jansen2009EnrichingSoftwareArchitecture}. In these other conceptualizations, architecture knowledge is explicit and often formal, and represented by different artifacts. These perspectives seek to explicitly define how architecture knowledge is created and used.

\subsection{Software Architecture Documentation}
\label{ss:SoftwareArchitectureDocumentation}

To be useful, software architecture knowledge must also be explicitly captured and represented as \textit{software architecture documentation}. Most frameworks for structuring and organizing software architecture documentation are based on the concept of \textit{architectural views}~\citep{KRU1995a}, which are different projections of knowledge about a system. Two popular frameworks for documenting software architecture are the Views and Beyond (V\&B) approach proposed by \cite{Clements:2010aa} and the approach of ~\cite{RW2012} (R\&W), based on the ISO/IEC 42010 standard. While both frameworks leverage the concept of architectural views, V\&B emphasizes the use of architecture styles while R\&W's focus is on stakeholders and their viewpoints. 

Despite the existence of well-supported frameworks for documenting software architecture, there exist many examples of documentation of the software architecture for a project that does not follow any pre-determined format, and simply presents and explains architectural decisions as a narrative complemented with informal diagrams~\citep{Brown:2012aa,SG2009a}. This type of documentation has been called \textit{essay-style documents (ESD)}~\citep{Robillard2016}, which can be contrasted with \textit{systematic approaches} such as V\&B or R\&W. 

The main distinguishing features between the ESD and systematic documentation styles are that the ESD conveys insights in the broader context of the project as elements in a sequential, almost story-like presentation, whereas the systematic styles advocate for a format where information can be retrieved as in a catalog. Because of their ready availability and accessibility, ESDs have been used as teaching resources for teaching software architecture~\citep{DAA2017a}.

Independently of the choice of representation, an open question in software architecture, and consequently for its documentation, is how much of it to produce~\citep{Fairbanks2010, Diaz2016}. Ultimately, how much effort to spend on architecture depends on the context~\citep{waterman2015}, and in particular the role it serves. With this study, we focus specifically on architectural documentation as a support for onboarding team members.

Relatively little empirical research has focused on assessing the impact of the way we represent software architecture knowledge. This is despite the fact that there are clear conceptual and structural differences among the recognized ways to document software architecture. de Graaf et al. conducted an experiment in an industrial context to assess whether architectural documentation organized as an ontology~\citep{GTLV2012} would be associated with increased efficiency and effectiveness of the retrieval of architectural knowledge~\citep{GLT2016a}, concluding that the ontology-based organization has the potential to improve the retrieval of architectural knowledge by better tailoring the documentation's organization to relevant questions. \cite{HKC2011a} and ~\cite{SHC2011a} investigated the respective merits of different combinations of textual and diagrammatic design information for software architecture. Their findings ``question the role of diagrams in software architecture documentation'' because of the comparative effectiveness of textual information. However, a more recent study by~\citet{JSD2020} came to the opposite conclusion. Based on a series of experiments with 240 software engineering students, the authors found that ``the graphical design description is better than the textual in promoting Active Discussion [sic] between developers and improving the Recall [sic] of design details''. In our study, we avoid the contention over the relative effectiveness of diagrams by employing documentation samples that include both text and diagrams.

\subsection{Onboarding in Software Projects}

Our research focuses on the role of documentation in supporting developers learning a software system new to them. This learning is typically done in the context of joining, or \textit{onboarding} a new software project. Researchers have investigated different aspect of onboarding, both in industry and in open-source communities.

An early study by \citet{DOB2010a} provided a grounded theory of onboarding based on the experience of 18 newcomers across 18 projects at IBM. The authors use a physical landscape as a metaphor for the environment that a software project represents. They report on the orientation aids and obstacles that newcomers encounter in a project landscape. In particular, this work underscores the importance of understanding a project's architecture when joining: ``Newcomers sought to understand the architecture mostly to get a broader view of their project, rather than because their tasks impacted or were directly impacted by the architecture. Without a good understanding of the architecture it was hard for newcomers to determine where their tasks fitted in the broader product and if their changes were complying with the existing architecture.''~\cite[p.278]{DOB2010a}

\citet{SGG2015a} later performed a systematic review of the literature on onboarding to identify the obstacles. These include, among the factors relevant to our study, too much documentation and software architecture complexity. Consequently, \citet{STG2018a} provide a list of guidelines for onboarding developers, which includes a recommendation about documentation, including with diagrams.

In a study in industry, \citet{BCS2018a} ``investigated the strategies employed by three different companies to onboard software developers in globally distributed legacy projects.'' The authors observed that one of the major onboarding challenges is, among others, ``the difficulty to learn the legacy code''. Follow-up work~\citep{BSD2020a} discusses the impact of formal training on onboarding performance. These observations serve to further motivate our study, as formal training in the context of onboarding requires the creation of documentation, whose design has potential alternatives.

Finally, we note that onboarding inevitably takes place in a given social and organizational context, and as such it is important to remember that non-technical forces, such as mentoring programs~\citep{FSB2014a,LH2015a} or prior socialization~\citep{CVD2015a}, can also impact the effectiveness of the onboarding process.

\section{Study Design}
\label{s:design}

Students enrolled in two different software architecture courses at two different institutions ({\A} and {\B}) had 75 minutes to read an architecture description of a software system, and answer different types of technical questions using a web form. The descriptions were of two types, and each description was randomly assigned to half the group. 
We analyzed the answers using a combination of qualitative coding and statistical analysis. 
The independent and dependent variables are defined in detail in Section~\ref{s:answerQuality}. 
The study design was approved by the ethics review board of both institutions. A complete replication package is available at \replPackage{}.

\subsection{Target System and Documentation}
 
As a target system, we chose {\tool} (Release 2.3), 
a software application for creating and editing diagrams in the Unified Modeling Language (UML) written in Java and developed by the second author. We chose this project because a credible architecture description (AD) can only reasonably be created by someone who has in-depth knowledge of the system. 

{\tool} provides core diagramming functionality complemented by the usual features of desktop applications, namely saving and opening files, undoing and redoing actions, copying and pasting, etc. The system release studied comprised approximately ten thousand lines of non-comment Java code (10 kSLOC) and came with a suite of 485 unit tests. 

The system's developer created two types of architectural descriptions that contained the same information. One description followed the highly structured style prescribed by the \textit{Views and Beyond} (V\&B) approach, and the other followed an essay style of documentation (ESD) (see Section~\ref{s:background}). The documents were in the English language.  
Because credible ADs also require reasonable knowledge of the AD format, and the first author was expert in V\&B, we chose the V\&B as the structured format for our study.

\begin{table}
    \centering
 \caption{Attributes of the Architecture Descriptions: Views \& Beyond (V\&B) and Essay-Style Document (ESD)}
    \begin{tabular}{lcc}
    \toprule
    & \textbf{V\&B} & \textbf{ESD} \\\midrule
Number of words & 789 & 993\\    
Number of diagrams (with legend) & 3 (1) & 3 (0)\\
Number of headers & 9 & 0\\
Number of lists (items) & 3 (13) & 0 \\
Number of links to external documents & 6 & 1\\
    \bottomrule
    \end{tabular}
   
    \label{tbl:docformat}
\end{table}

Table~\ref{tbl:docformat} contrasts the salient differences between the two documents. We took care to ensure that both versions of the architecture description conveyed the same information. The ESD document is slightly longer (in number of words) to account for the narrative style. Both ADs include three diagrams: a UML Sequence diagram, a UML Class diagram, and a UML Package diagram.
Our replication package includes the full documents used including all diagrams.
These diagrams were identical in both versions of the documents except for the presence of a legend in one diagram of the V\&B document. All differences are thus structural, with the consistent theme that the V\&B document is more structured and the ESD is more narrative. 

For example, the V\&B document partitions the text using nine standardized section headers, such as ``Primary Presentation'' and ``Element Catalog'', while the ESD organizes the information into coherent fragments using paragraphs, but without using headers to break the text. Thirteen pieces of information are provided as bullet list items in the V\&B documents, while the information is in-line in the ESD. Finally, although both documents mention the same Application Programming Interface (API) elements (e.g., \textsf{Point}, \textsf{BorderPane}), the mentions in the V\&B's element catalog are linked to the corresponding API documentation, whereas those in the ESD are simply denoted with a fixed-width font. 

\subsection{Questionnaire}
\label{s:questionnaire}

The questionnaire we used as a research instrument was structured in three parts: \textit{background}, \textit{architecture understanding questions}, and \textit{subjective experience}. Because we collected the data anonymously, we also required the participants to explicitly state their institution ({\A} or {\B}) and the documentation format randomly assigned to them (ESD or V\&B), as well as consent to the collection of their data. Respondents were neither prohibited from using nor encouraged to use other sources to answer the questionnaire, including the project's source code repository. 

\paragraph{Background}

\begin{table}
\centering
\caption{Background Questions (summarized). An asterisk (*) indicates that multiple answers are possible.}
    \begin{tabular}{lp{10cm}}
    \toprule    
    B1. & What is your reading proficiency in English?  \newline (fluent/modest/limited)\\
    B2. & Have you taken any high-level relevant courses?  \newline (yes/no)\\
    B3. & How long have you been programming? \newline ($<$1/1-2/3-5/5+ years)\\
    B4*. & What software development experience do you have? \newline  (coursework only/internships/professional work outside school)\\
    B5*. & What programming languages are you comfortable with?  \newline (C++/C/Java/Python/JS/C\#/Other)\\
    B6. & What is your level of experience with UML?  \newline (never/courses/read diagrams/created diagrams)\\
    B7*. & What is your previous exposure to {\tool}? \newline  (never heard/used once/used frequently/looked at source code)\\
    \bottomrule
    \end{tabular}
    \label{tbl:background}
\end{table}

The questionnaire included seven questions aimed at better understanding the background of the respondents (Table~\ref{tbl:background}), including English proficiency (B1), formal education on software design and architecture (B2), and technical background and experience (B3--B6). We also asked about prior familiarity with {\tool}.

\paragraph{Architecture Understanding Questions}

The second part of the questionnaire consisted of six two-part questions which asked respondents to read the documentation and answer a particular question about the system. 
For each of these questions, we asked respondents to explain the process used to obtain the answer with a follow up question, the same for all questions. 
The questions followed a progression in their level of difficulty and learning objective according to Bloom's revised taxonomy~\citep{Bloom2001}.

Table~\ref{tbl:questions} describes our architecture understanding questions.\footnote{The last two questions exhibited limited completion rates because of time constraints. The table only includes the four questions we analyzed.}  
Each question maps to a learning objective from Bloom's revised taxonomy that ranges from Level I~(Remembering) to Level VI~(Creating)~\citep{Bloom2001}. The last row of the table provides the text of the general follow-up question we applied to each question. 

Specifically, Question 1 asked participants to study the architecture and related resources and \textit{find}, \textit{choose}, or \textit{select} a relevant module from a list of choices (the italicized terms are from Bloom's revised taxonomy's action verbs for the corresponding level). Question 2 asked participants to identify the service layer in the architecture, \textit{infer} the services it provides, and \textit{rephrase} and \textit{summarize} them. Question 3 asked participants to \textit{identify} a pattern in the design of the application and to \textit{apply} this pattern to \textit{construct} a solution. Finally, Question 4 asked participants to \textit{elaborate} an \textit{original} \textit{solution} by \textit{adapting} the existing architecture and \textit{designing} new functionality.

\begin{table*}
\caption{Architecture Understanding Questions. Questions 1--4 each had a common second part, G (for ``General''). For each question numbered 1--4 and G, the table includes the exact text of the question
and the corresponding learning objective level from Bloom's revised taxonomy~\citep{Bloom2001}. The answer format for Question 1 was multiple-choice, and all others were free-from text.} 
    \begin{tabular}{cp{7.8cm}c}
    \toprule    
    \textbf{Q\#}  & \textbf{Text} & \textbf{Bloom}\\\midrule
    1 & Which module contains the code for starting the application? & I.--Remembering\\
    2 & What are the architectural responsibilities of the Services layer? & II.--Understanding\\ 
    3 & How would one add a new diagram type to {\tool}? & III.--Applying\\
    4 & How would you add a status bar to the area that shows a diagram? & VI.--Creating\\ \midrule
    G & How did you determine the answer to the previous question? Please describe which document, on-line resource, source code comment, or other artifact you have used to answer...
    & N/A\\
       \bottomrule
    \end{tabular}
    \label{tbl:questions}
\end{table*}

\paragraph{Subjective Experience} The third part of the questionnaire asked respondents to reflect on their experience using the architecture documentation and on what they had learned from the session (Table~\ref{tbl:experience}).

\begin{table}
\caption{Subjective Experience Questions (summarized). Answers were ordered choices from Strongly Disagree to Strongly Agree.}
    \begin{tabular}{ll}
    \toprule    
    SE1. & The document was easy to navigate.\\
    SE2. & I would need assistance to use this document.\\
    SE3. & There was too much inconsistency in the  document.\\
    SE4. & I needed to learn a lot before I could get going.\\
    SE5. & This document gave me a sense for {\tool}'s vision.\\
    SE6. & Readers would learn to use this document very quickly.\\
    SE7. & I could see using this document in practice.\\
    SE8. & What is your impression of the document you consulted.\\
    \bottomrule
    \end{tabular}
    \label{tbl:experience}
\end{table}

\begin{table*}[h]
\caption{Background Characteristics of the Participants}
    \begin{tabular}{lp{3.2cm}p{7cm}}
    \toprule    
    \textbf{Q\#}  & \textbf{Mnemonic} & \textbf{Description} \\ \midrule
    BC1. & English language & \textbf{56} (85\%) Fluent vs. \textbf{9} (14\%) Modest \\
    BC2. & Software design courses & \textbf{56} (85\%) Yes vs. \textbf{9} (14\%) No\\
    BC3. & Programming  & \textbf{37} (57\%) 3--5 years; \textbf{23} (35\%) $>$ 5 years; \textbf{5} (8\%) 1--2 years \\
    BC4. & Work experience & \textbf{38} (58\%) interns; \textbf{18} (28\%) professional developers; \textbf{19} (29\%) no work experience.\\
    BC5. & Programming languages & \textbf{60} (92\%) indicated familiarity with Java\\
    BC6. & Experience with UML & \textbf{13} (20\%) in industry; \textbf{46} (71\%) in courses; \textbf{6} (9\%) none\\
    BC7. & What is your previous exposure to {\tool}? & See Figure~\ref{fig:prev_tool}.\\
    \bottomrule
    \end{tabular}
    \label{tbl:background2}
\end{table*}

\subsection{Participants} 
\label{ss:Participants}

We recruited students in our two software design and architecture courses available to both senior undergraduate students and graduate students. The courses were offered at {\A} and {\B} simultaneously in the Winter 2019 semester. We received 65 responses to the questionnaire, 25 from {\A} (86\% response, i.e., consented to inclusion in the study), and 40 from {\B} (80\% response). In both course sections the students received participation marks for taking part in the study, irrespective of whether they consented to our use of their data or not.
A total of 29 respondents used the ESD type document, and the remainder (36) the V\&B document.\footnote{The uneven split is a consequence of the fact that we could not determine in advance who would consent to participate in the study.}

Table~\ref{tbl:background2} summarizes the background of our participants. These were senior students, so nearly all were familiar with Java (the implementation language of {\tool}) and most had done additional software design courses and multiple years of programming experience. In addition, 71\% had industry experience. The majority of students (85\%) indicated fluency in reading English, and the remainder indicated modest proficiency. Only six participants indicated they had never used UML; the remainder had at least used it in course work, and 13 had used it in industry. Five students had less than three years of programming experience, while the remainder had three or more. 

\begin{figure}[hbt]
\centering
  \includegraphics[width=0.7\textwidth]{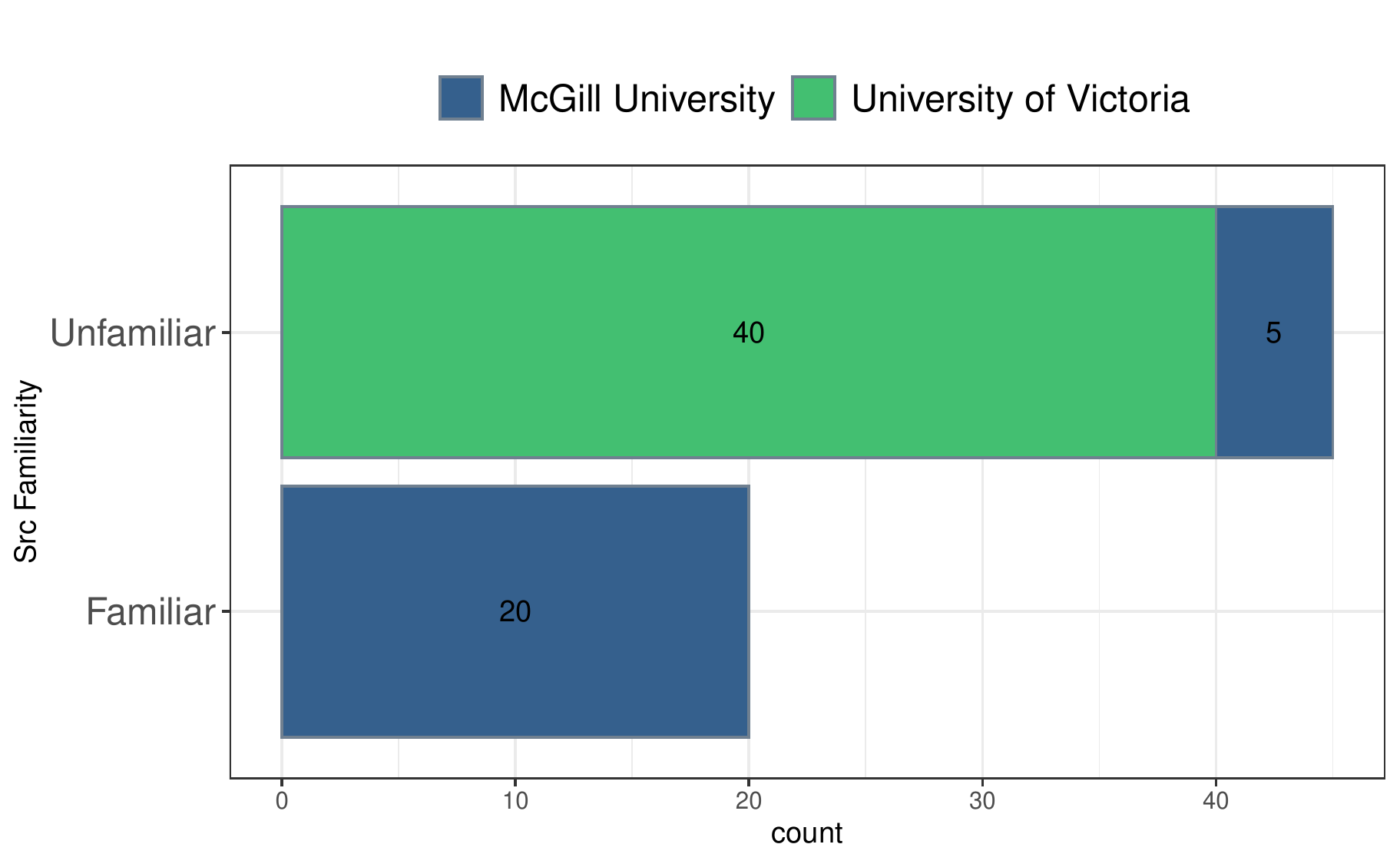}
  \caption{Familiarity with {\tool} source code by institution.}
  \label{fig:prev_tool}
\end{figure}

There was a significant difference in familiarity with the subject system {\tool}, as shown in Figure \ref{fig:prev_tool}. Students at {\A} were mostly familiar with it (they had previously used the tool and/or looked at its source code) and students at {\B} were unfamiliar with it. This dichotomy was expected given that {\tool} is developed at {\A} and students are exposed to it as part of undergraduate classes. Nevertheless, no participant had contributed to the software, had knowledge of its overall architecture, or had investigated any of the questions in our questionnaire.

The dichotomy in participant familiarity enables an important aspect of the study, namely to include prior familiarity with a project as an independent variable. For the purpose of the analysis, we categorized respondents who indicated ``I have never heard of {\tool}'' or ``I have used {\tool} once or twice'' to be \textit{unfamiliar} with {\tool} and respondents who indicated ``I have looked at {\tool} source code and/or documentation before this'' to be \textit{familiar}.
 
\section{Analysis and Findings}
\label{sec:findings}

Our first step was to evaluate the answers to the architecture understanding questions (Section~\ref{s:answerQuality}). We then posited different statistical models that incorporate the independent variables \textit{document type} and \textit{prior knowledge}, as well as other potential causal factors. We subsequently assessed which statistical model was most explanatory of higher quality answers. For acceptable answers, we then investigated the process followed by the participant to answer the question, and specifically to what extent that process involved perusing the documentation, once again contrasting between the two different documentation formats (Section~\ref{s:usageProcess}). 
Finally, we analyzed the responses to the \textit{subjective experience} part of the questionnaire (Table~\ref{tbl:experience}) to learn from richer data how the participants reacted to the two different document types (Section~\ref{s:Experience}). 

\subsection{Analysis of Answer Quality}
\label{s:answerQuality}

\begin{figure}\centering
    \includegraphics[width=0.8\linewidth]{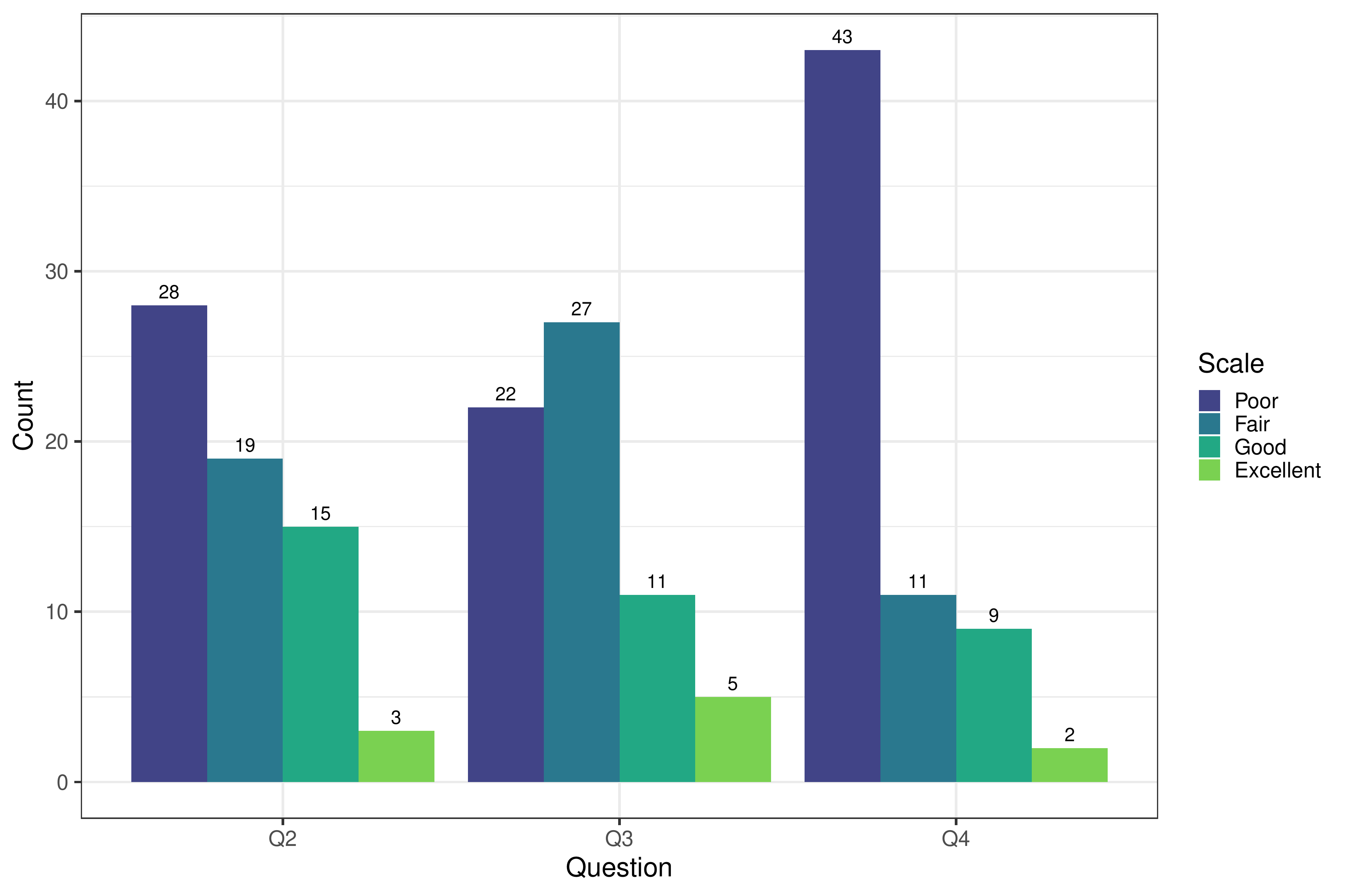}
    \caption{Distribution of Answer Levels. N=65.}
    \label{fig:useful_dist}
\end{figure}

Question 1 was a multiple-choice question with an objective outcome. For Questions 2--4, the second author reviewed each answer in scored it on the following ordinal scale:
\begin{itemize}
	\item Level 0: (Poor): Incorrect or overly vague. Demonstrates no actionable knowledge of the architecture.
	\item Level 1: (Fair): Minimal, incomplete, somewhat vague, or repeats sentences of the document verbatim. Demonstrates only the shallowest possible knowledge of the architecture.
	\item Level 2: (Good): Demonstrates a basic level of interpretation. Knowledge of the architecture that could be usable in practice.
	\item Level 3: (Excellent): Demonstrates a clear understanding of the architecture as well as relevant implementation details.
\end{itemize} 

Although the procedure involves personal judgement, it was not possible to perform an independent evaluation with inter-reviewer reliability measurement because evaluating architectural insights on a non-trivial software system requires a great deal of expertise with the system, and only one researcher had this expertise (see Section~\ref{s:threats} for additional details on how we mitigated the threat of bias).

We focus on questions 2--4 as the finer-grained evaluation scale and open answer format provide richer insights and avoid the threat of respondents answering correctly by chance.
Figure \ref{fig:useful_dist} shows the distributions of levels.  As can be seen from the figure, many of the answers are poor. This result can be readily explained by the combination of two factors: a) understanding software architecture is challenging~\citep{DOB2010a}, and b) successfully overcoming this challenge in an experimental setting requires a minimum of motivation. Without a means to measure motivation, we must assume that unmotivated participants were distributed randomly. For the analysis of information sources (Section~\ref{s:usageProcess}), where we must interpret textual answer, we mitigate the threat by only considering questions with a score of Fair or better.

We created inferential models to understand the factors associated with higher scores. Our models are Bayesian ordered categorical regression models~\citep{Brkner2019} and model the ordinal question score as a factor of two main predictors. One is whether the architecture description format, namely Views and Beyond (V\&B) or Essay Specific Documentation (ESD). The second is previous familiarity with the \tool{} source code. 

\subsubsection{Causal Models}
\begin{figure}
    \includegraphics[width=\textwidth]{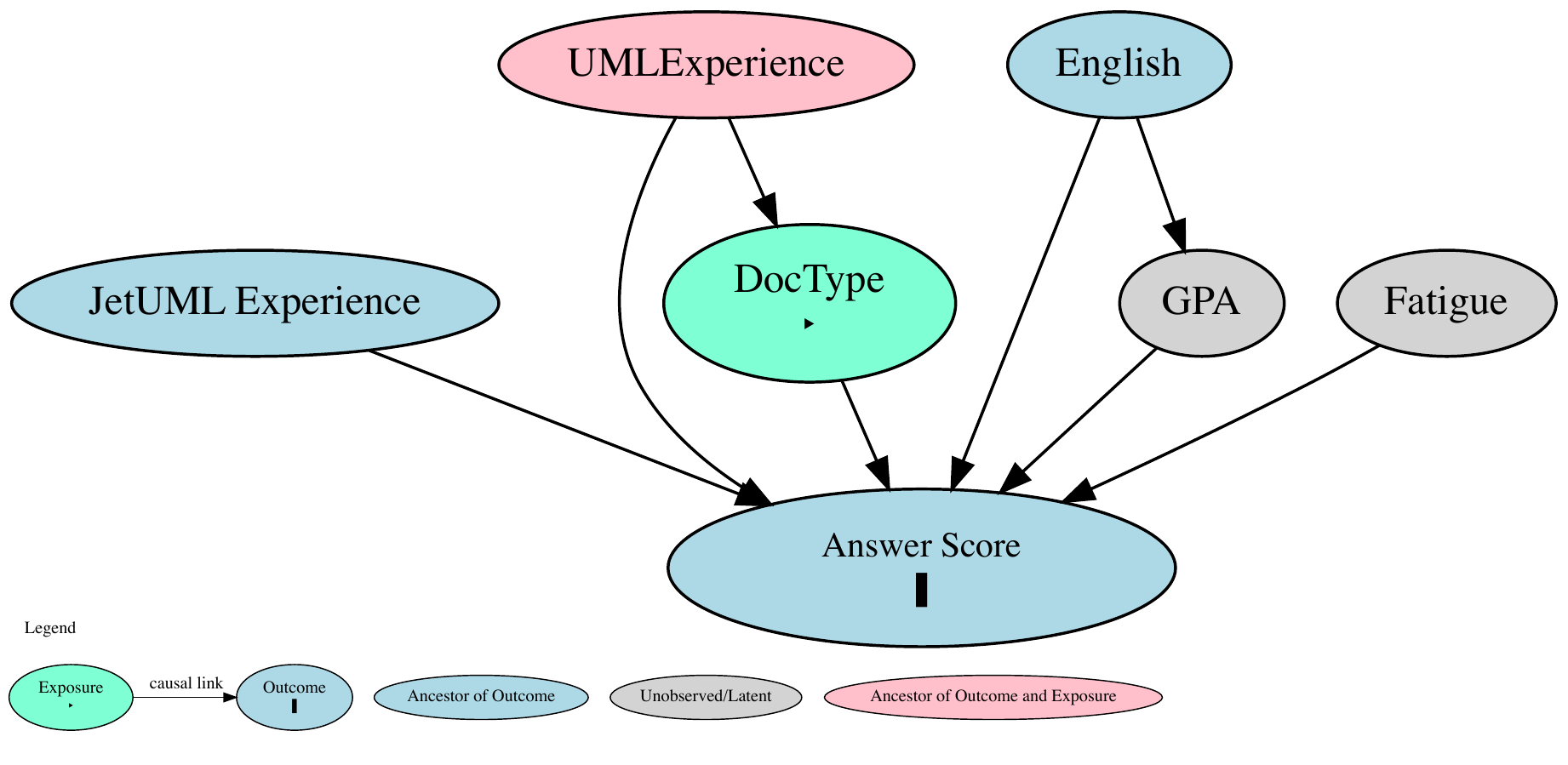}
    \caption{Potential causal graph showing predictors (blue), response (blue with thick bar), the treatment (\textsf{DocType}, which we controlled) and unobserved/latent variables (grey). This graph represents Model 1. \textsf{UMLExperience} in red reflects that it may be a confound of \textsf{Answer Score}, since \textsf{UMLExperience} may influence one's comprehension of the document type.}
    \label{fig:causal}
\end{figure}

Figure \ref{fig:causal} represents the analysis question as one possible causal model \citep{Rohrer2018}. 
Our ordered categorical regression is based on this causal graph. 
The causal model is a Directed Acylic Graph (DAG) in which directionality of links reflects a causal/influence relationship between variables. For example, the graph reflects our (a priori) belief that \textsf{DocType} (i.e., V\&B or ESD) causes or changes the \textsf{Answer Score}. Since we varied the \textsf{DocType} it is a \textit{treatment or exposure} variable. We also measured (but did not control) \textsf{\tool\textsf{Experience}}.

There are a number of latent variables we might represent and also some explicit constructs we measure (\textsf{English} facility, \textsf{UML Experience}). Together those factors will cause the predicted score for the answer score. 
Some latent factors might include a) \textsf{Fatigue}, or how tired the student is (maybe they did not feel like answering, and were not forced to), which might be of growing importance in later questions,
b) how capable a student they are (e.g., as approximated by \textsf{GPA}), and other factors including gender and age. 

Our independent variables, derived from Table \ref{tbl:background2}, include previous familiarity with the source code of JetUML (\textsf{JetUML Experience}), the type of document they were using (\textsf{DocType}), previous software design courses \textsf{(DesignCourse)}, English language familiarity (\textsf{English}), \textsf{GenExp} (previous industry work), and amount of previous programming experience (\textsf{ProgExp}). Our dependent variable is the score of the answer (\textsf{Answer Score}) from 0 to 3 as mentioned above. We examine these models on all three of the architecture understanding questions. 

A causal graph can capture the relationship between the variables. Fig. \ref{fig:causal} shows an example that represents $\mathcal{M}_{1}$ from Table \ref{tab:models}. 
The goal of the analysis is to find a model that best explains the data with the fewest predictors. 

Many models are possible; not least, the combination of all possible predictors. This raises the question about which, and how many, predictors to include in the model~\cite[Section 7.5]{mcelreath20}.
As an exploratory study, we focus on four models that represent different potential causal attributes of our problem. 
Other combinations are possible, but we focus on these four as they represent the most plausible causal relationships.
Adding more predictors to the model raises the risk of overfitting (biasing) to the collected data and reduces our ability to explain more general phenomena.
We then compared the explanatory power of our different models. The more a model can explain the data, the more indicative that is of potentially important effects. 

\begin{table}
	\caption{Possible explanatory models for the \textsf{Answer Score} of an architecture answer for Questions 2, 3, and 4. Tilde notation should be read as ``dependent variable (left side) is explained by independent variable interactions" (right side).}

\begin{tabular}{cp{0.7\linewidth}}
\toprule
Model Number & Statistical Model \\
\midrule
	$\mathcal{M}_1$ & \textsf{Answer Score $\sim$ DocType + JetUML Experience + UMLExperience + English }\\
	$\mathcal{M}_2$ & \textsf{Answer Score $\sim$ DocType + JetUML Experience + ProgExp + DesignCourse + ProgLang} \\
	$\mathcal{M}_3$ & \textsf{Answer Score $\sim$ DocType + ProgExp + DesignCourse + JetUML Experience + ProgLang + GenExp + UMLExperience + English} \\
    $\mathcal{M}_4$ & \textsf{Answer Score $\sim$ DocType + JetUML Experience  + GenExp } \\
	\bottomrule
\end{tabular}
	\label{tab:models}
\end{table}

\subsubsection{Bayesian Ordered Categorical Regression}
Ordered categorical regression models the data responses (here, the architecture question score $Q$) as being driven by an underlying latent variable $\tilde{Q}$ which is divided into the different score categories (i.e., $<1,1-2,2-3,3+$) with cutpoints. The cutpoints determine, given the inferred result of $\tilde{Q}$ for an individual response, which ordinal category to assign. See \cite{Brkner2019} whose method we follow in this analysis for more details. 
The key insight is that using cutpoints properly treats the data as ordered adjacent categories rather than a continuous response.

\subsubsection{Model Comparison}
We fit the four models based on the different combinations of predictors. Our replication package contains the code to reproduce these statistical analyses.
We assess which of the four models best explains the data. This is a Bayesian workflow as described in \cite{GelmanWorkflow}, and elaborated for software research by \cite{Torkar2021}.
A model comparison approach focuses on building an adequate explanation for the observed data that can be useful in answering questions, doing decision analysis, or making predictions. 
It does not imply it is the \emph{best} possible model, just one that, given the various factors in our causal model, best explains (`is least surprised by') the data.
Model comparison allows us to evaluate which model is best at explaining the observed data. We do this with the leave-one-out cross-validation approach (LOO) described in \cite{Vehtari2016}.

We compare our four different predictive models for each question and choose the one that is most informative using Leave One Out sampling (e.g., comparing the model trained on 64 datapoints against a held out single data point, 65 times). 
We repeat the process for each of the three questions. 
The outcome from LOO is a model comparison value based on the differences in Expected Log Predictive Density (dELPD)\citep{Vehtari2016PracticalBayesianModel}. 
The best model has a dELPD of 0, and the other models are compared to that result. The difference reports that the model is a better fit for the data, i.e., achieves a bias-variance trade-off better than the alternative models~\citep{Furia2022}.

Where dELPD scores are less than an integer multiple of the standard error of that difference (dSE), as in this case, other factors such as our domain knowledge (i.e., how likely is a predictor to influence the result) and model parsimony (fewer predictors are preferred over more) are also important. We are likely seeing the effects of noisy data and low sample size.
Table \ref{tab:looic} shows the results. 
Since the values of dSE include 0 (e.g., for Q2, the difference between M2 and M1 is -0.1, but the SE is 2.8), there is no significant difference between models. We choose model 1 as best combining simplicity (in numbers of variables) and accuracy, of the four. 
That is, \textsf{Answer Score $\sim$ DocType + JetUML Experience + UMLExperience + English}  best explains  the observed answer scores.

\begin{table}
    \centering
    \caption[]{Comparing models using Leave One Out information criteria. dELPD = difference in Expected Log Predictive Density, a measure of predictive accuracy within-sample, compared with the best model. dSE = the  standard error of this difference in predictive density}.
\begin{tabular}{cccc}
    \toprule
    Question & Model & dELPD & dSE \\
    \midrule
    \multirow{3}{*}{Q2} & $\mathcal{M}_2$ & 0.0  & 0.0 \\
                        & $\mathcal{M}_1$ & -0.1 & 2.8   \\
                        & $\mathcal{M}_4$ & -0.2 & 2.5   \\
                        & $\mathcal{M}_3$ & -3.0 & 1.8 \\
    \midrule
    \multirow{3}{*}{Q3} & $\mathcal{M}_4$ & 0.0 & 0 .0  \\
                        & $\mathcal{M}_1$ & -0.1 & 1.6 \\
                        & $\mathcal{M}_2$ & -0.1 & 2.5 \\
                        & $\mathcal{M}_3$ & -0.3 & 3.3 \\
    \midrule
    \multirow{3}{*}{Q4} & $\mathcal{M}_1$ & 0.0 & 0.0\\
                        & $\mathcal{M}_4$ &  -2.5  & 6.5  \\
                        & $\mathcal{M}_2$ & -3.3  & 2.3 \\
                        & $\mathcal{M}_3$ & -4.6  & 1.3 \\
    \bottomrule
\end{tabular}
\label{tab:looic}
\end{table}

\subsubsection*{Model Results}

We use $\mathcal{M}_1$ to analyse the coefficients for the regression models. Coefficients represent the additive change in the score caused by that coefficient. The inference procedure generates a coefficient and associated credible interval (the Bayesian analogue for a confidence interval~\citep{Torkar2021}).

Tables \ref{tab:q2_coeff}, \ref{tab:q3_coeff}, and \ref{tab:q4_coeff} report regression coefficients for the three questions. We broadly conclude that a coefficient has inferential value if its 95\% credible interval (CI \textbf{u}pper and \textbf{l}ower) does not include 0, which is indicative of a null effect. We have bolded those cases in the tables. 

We do not see consistency from question to question for our predictors, except with the predictor \textsf{JetUMLYes}, reflecting prior JetUML experience. In Q2 we see for example some effect from English facility, but that disappears in Q3 and Q4. Our conclusion is therefore that across Q2, Q3, Q4 only JetUML experience is a substantial influence on the answer score. For example, in Q2 (Table \ref{tab:q2_coeff}) it seems to add nearly a whole point (i.e., 25\%) to the rating. 
\textsf{DocType} does not have influence except in Q3, where using the V\&B approach reduces the score.
 
\begin{table}
    \centering
    \caption{Coefficients for Q2, Model 1. Bold coefficients reflect a predictor where the 95\% credible interval (the Bayesian form of a confidence interval) does not include 0. 
    Predictors: DoctypeTypeV = respondent used V\&B; JetUMLYes = respondent had familiarity with JetUML source code; UMLExpNo = respondent had no previous UML experience.
    }
    \begin{tabular}[]{cccc}
        \toprule 
        Value & Estimate & l-95\% CI & u-95\% CI \\
        \midrule
        DoctypeTypeV &  -0.38 & -1.05 &  0.27 \\
        JetUMLYes  &  \textbf{0.95} &  0.28 &  1.68 \\
        UMLExpNo   &  -0.79 & -2.33 &  0.46 \\
        EnglishModest & \textbf{-1.71} & -3.61 & -0.28 \\
        \bottomrule
    \end{tabular}
\label{tab:q2_coeff}
\end{table}

\begin{table}
    \centering
    \caption{Coefficients for Q3, Model 1}
    \begin{tabular}[]{cccc}
        \toprule 
        Value & Estimate & l-95\% CI & u-95\% CI \\
        \midrule
        DoctypeTypeV &  \textbf{-1.03} & -1.78 & -0.30 \\
        JetUMLYes &      \textbf{1.02} &  0.33 &  1.75 \\
        UMLExpNo &      -1.12 & -2.66 &  0.19 \\
        EnglishModest &  0.25 & -0.80 &  1.25 \\
        \bottomrule
    \end{tabular}
\label{tab:q3_coeff}
\end{table}

\begin{table}
    \centering
    \caption{Coefficients for Q4, Model 1}
    \begin{tabular}[]{cccc}
        \toprule 
        Value & Estimate & l-95\% CI & u-95\% CI \\
        \midrule
        DoctypeTypeV &  -1.04 & -1.84 & -0.31 \\
        JetUMLYes &      \textbf{1.03} &  0.30 &  1.83 \\
        UMLExpNo &      -1.15 & -2.78 &  0.20 \\
        EnglishModest &  0.25 & -0.76 &  1.25 \\
        \bottomrule
    \end{tabular}
\label{tab:q4_coeff}
\end{table}

\begin{figure}
    \includegraphics[width=\linewidth]{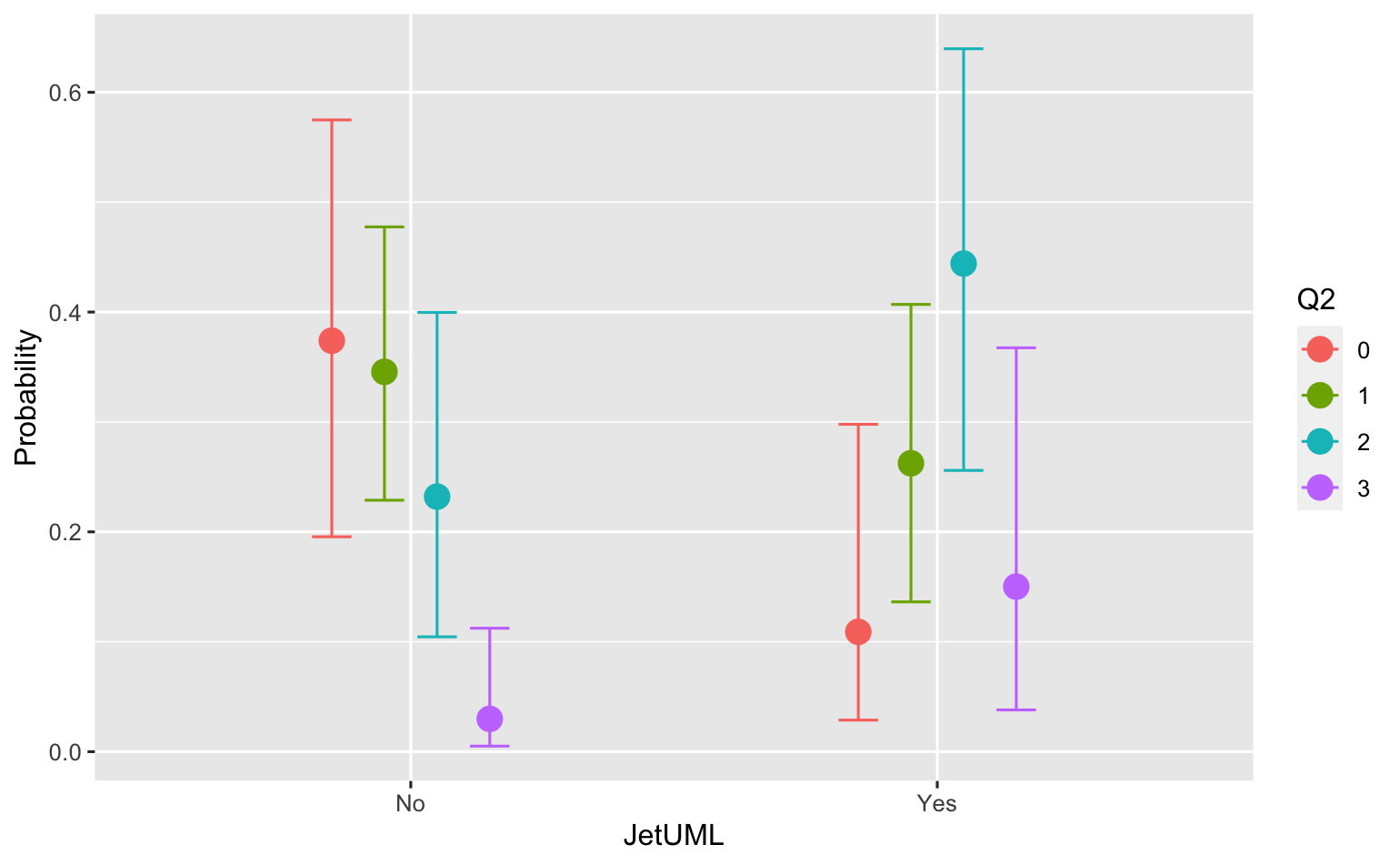}
    \caption{Marginals for JetUML predicting answer rating. Two columns indicate respondents who had (Yes) or did not have (No) prior JetUML knowledge. The y-axis reflects the probability the predictor influenced the score (dots with 95\% credible intervals). From left to right, red = score of 0, green = score of 1, blue = score of 2, purple = score of 3.}
    \label{fig:marginal_q2_JetUML}
\end{figure}

A strength of Bayesian inference is the ability to leverage the posterior distribution for further analysis. 
We can look at the marginal effects of a given predictor on the outcomes, as shown in Figure \ref{fig:marginal_q2_JetUML}. 
The figure captures the impact of previous JetUML knowledge on the usefulness of the answer. We can see that not knowing JetUML (left side of Fig. \ref{fig:marginal_q2_JetUML}) results in a much lower probability of getting a score of 3 (the highest), by examining the rightmost purple dot and credible interval. Similarly, not having previously been exposed to the code of JetUML (left side) meant a marginal probability of nearly 40\% for scoring a 0, as opposed to those who were familiar with the code of  JetUML, with a probability of 12\% for scoring 0. This accords with Table \ref{tab:q2_coeff}. There, JetUML knowledge accounts for an increase of 0.95 [95\% CI 0.28,1.68] in score. 

\begin{myboxi}[Observation 1: We observed no significant impact of the documentation format on the outcome of software architecture understanding tasks. Prior exposure to the source code of the system was the dominant factor affecting the performance of the software architecture understanding tasks.] 
Software architecture documentation is a feature of a software project's landscape that can help orient new project contributors, but there exist other, potentially more effective orientation aids, such as code-oriented tasks and code walkthroughs~\cite[Table~2]{DOB2010a}. Our first observation suggests that integrating code-oriented tasks as part of onboarding practices may have more impact than the choice of documentation format.
\end{myboxi}

\subsection{Analysis of Information Sources}
\label{s:usageProcess}

To answer questions, participants were directed to look at the architectural documentation, but also had access to the source code of the system, and were free to search the Internet as they pleased. We investigated whether the format of their architectural description was associated with the use of different information sources for producing an answer.

For answers we scored as \textit{Fair} or better (i.e., Levels 1--3), we analyzed the answer to the corresponding process question (G in Table~\ref{tbl:questions}) to determine the type of information the participants used to produce their answer. As determining this information involved almost no subjectivity, we reviewed and coded all answers collaboratively in a single session using the three codes: \textit{Source Code}, \textit{Description}, and \textit{External}. An answer could receive any subset of the three codes, as applicable. 

\begin{table}
\centering
\caption{Sources of information for \emph{Fair} or better answers. N=102. Zero or more sources were possible.} %
    \begin{tabular}{cccccc}
    \toprule
    \textbf{Q\#} & \textbf{Docs} & \textbf{Code}  & \textbf{External} & \textbf{No Source} & \textbf{Total \emph{Fair} or better} \\\midrule
    2      &   32      & 4          & 1    &   5    & 37         \\
    3      &   22     & 17          & 0    &   12    & 43      \\ 
    4      &   13     & 6          & 2     &   7    & 22      \\  
    \bottomrule
    \end{tabular}
    
    \label{tbl:other_src}
\end{table}
Table \ref{tbl:other_src} shows a breakdown by information location type. We observe that for the simpler question (2), most answers were produced from the documentation only. For the questions with higher-level learning objectives (3 and 4), a greater number rely on source code. We confirm this association with a contingency table to which we applied Fisher's exact test (Table~\ref{t:contingencyTable}).

\begin{table}
\centering
\caption{\label{t:contingencyTable}Number of \textit{Fair} or better answers to Q2--Q4, that leveraged perusal of the \tool{} source code (`Code'), vs. in the provided documentation or external references (`Doc/Ext'), broken down by document type. Use of both Code and Doc/Ext was possible. }
\begin{tabular}{rrrrrr} \toprule
    & \multicolumn{2}{c}{\textbf{ESD}} &  \multicolumn{2}{c}{\textbf{V\&B}} & \\ \midrule
      & Code & Doc/Ext & Code & Doc/Ext & Total\\ 
      Question 2, Bloom Level II            & 1  & 17 & 3  & 15 & 36\\
      Questions 3 \& 4, Bloom Level III, IV & 12 & 19 & 11 & 17 & 59 \\ \midrule
      Total                                 & 13 & 36 & 14 & 32 & 95 \\ \bottomrule
\end{tabular}
\end{table}

Answers to questions that require applying and creating activities (3 and 4) are statistically significantly associated with the use of the source code in the information search process ($p$-value 0.0046, odds ratio 5.05). 
For completeness we investigated the relation between a) documentation format and code familiarity, as well as b) between answer sources and either documentation format or prior familiarity. We did not detect any significant association at the $\alpha=0.05$ level (e.g., Table \ref{tab:fisher_docstyle}).

\begin{myboxi}[Observation 2: Answers to questions that required \textit{applying} and \textit{creating} activities were statistically significantly associated with the use of the system's source code to answer the question whereas the document format or level of familiarity with the system were not.] 
We hypothesise that our architecture documents did not provide sufficient details to properly support design tasks. 
Our experimental artifacts were specifically created to represent typical software architecture documents.
Such documents may not be the best medium to support design reasoning. 
This is because an architecture document's level of abstraction may not support the detail needed to begin implementing application features. 
This conclusion is consistent with the findings of~\cite{DOB2010a}, that ``Participants said they learned more efficiently by experimenting with the code than
by attending courses or reading documentation.''
\end{myboxi}

\begin{table}[ht]
\centering
\caption{Relationship between documentation style and code familiarity for answers scored `Fair' or higher for Q2--Q4 (N=102). Fisher's exact test shows differences between cells are not unusually distributed at $\alpha = .05$.}
\begin{tabular}{rrr}
  \toprule
 & ESD & V\&B \\ 
  \midrule
Familiar &  21 &  31 \\ 
  Unfamiliar &  30 &  20 \\ 
   \bottomrule
\end{tabular}
\label{tab:fisher_docstyle}
\end{table}

\subsection{Analysis of Participants' Experience}
\label{s:Experience}

\begin{figure}[bt]
  \includegraphics[width=0.9\textwidth]{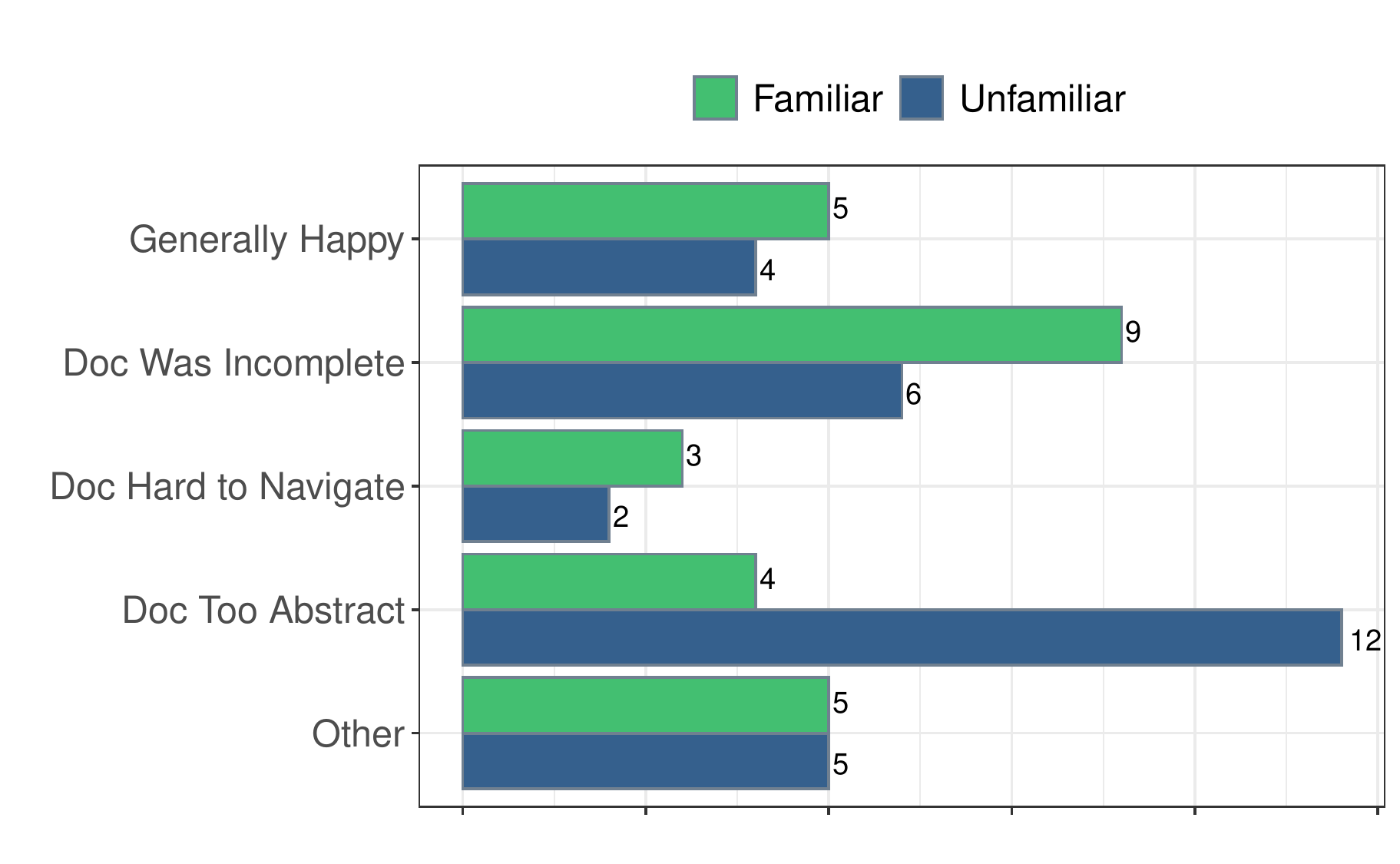}
  \caption{Respondent impressions of the documentation based on familiarity with the underlying tool. `Other' aggregates 10 codes with a frequency of only one or two.}
  \label{fig:subj_code}
\end{figure}

The last section of the questionnaire asked respondents to ``share your general impression of the effectiveness of the document you consulted.'' 
We analyzed the responses to this question qualitatively using a closed set of codes. Figure~\ref{fig:subj_code} shows the aggregated results, blocked by respondent familiarity. Respondents unfamiliar with \tool{} found the architecture documentation too abstract (3x as likely as those with familiarity), while those with familiarity were more likely to mention incompleteness (1.5x more likely). 

 \begin{figure*}[bt]
  \includegraphics[width=0.9\textwidth]{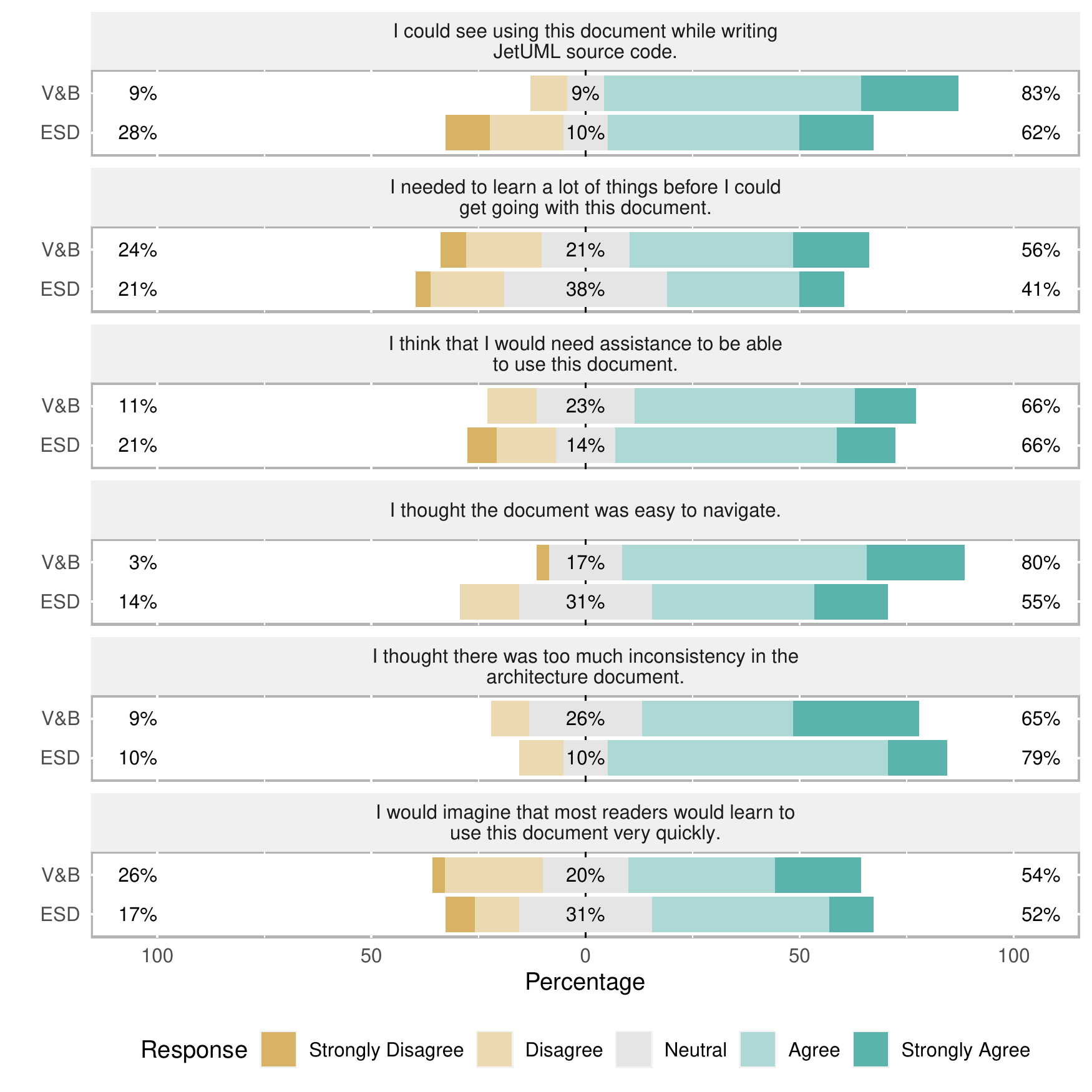}
  \caption{Likert-scale assessments of the two architectural descriptions. Percentages refer (from Left to Right) to proportions who Strongly Disagree/Disagree (unfavourable), Neither, and Agree/Strongly Agree (favourable). }
  \label{fig:likert}
\end{figure*}

We also asked a series of questions using a five-level Likert scale. 
The scale ranged from ``Strongly Disagree" to ``Strongly Agree". Figure \ref{fig:likert} captures the different assessments. The questions exhibiting the starkest differences between ESD and V\&B are \emph{Navigability} and \emph{Helpfulness for coding}. For \emph{Navigability} (``I thought the document was easy to navigate''), ESD has 55\% favourable versus V\&B with 80\% favourable. For \emph{Helpfulness} (``I could see using this document while writing {\tool} code``, top row in Figure \ref{fig:likert}), ESD was more negatively perceived, with  28\% disagreeing or strongly disagreeing with the sentiment, while V\&B had 83\% agreeing or strongly agreeing it would be useful for writing code. 

We used a similar cumulative Bayesian ordinal model~\citep{Brkner2019} to assess the differences between ESD and V\&B documents for these two questions. Even though there seems to be a difference in perception in favour of V\&B, we could not validate the significance of this difference statistically.

\begin{myboxi}[Observation 3: Subjective sentiment about the documentation format was comparable for the V\&B and ESD formats we used in the study. More participants agreed that the V\&B document was easier to navigate and use for writing code{,} but this relation was not statistically significant.]
    A single architectural documentation format may not suitably support project newcomers both for getting a broader view of their projects and for design tasks. The survey results indicate that the ESD format may be better suited for overviews and the V\&B format for coding tasks. The implication is to select a documentation format with not only the audience in mind, but also the specific usage scenario for the architecture documentation.
\end{myboxi}

\subsection{Threats to Validity}
\label{s:threats}

\subsubsection*{Construct Validity} The  general construct of interest is \textit{software architecture understanding}, which we study via the more specific construct of \textit{ability to answer questions about the architecture of a software system.} We measured this latter construct subjectively using a four-level scale (0-3). We used this simple ordered rating to avoid the imprecision associated with a finer-grained scale. At the same time, the measure loses fidelity, as insightful answers are considered similar to those providing less insightful information. 

Investigator bias is a threat given that an author of the paper scored the answers.  
To help prevent bias, we concealed the responses to the associated independent variables of \textsf{DocType} and \textsf{JetUML Experience} while assessing answer usefulness.
For transparency we include the full set of rated answers as part of our replication package. 

Our architectural descriptions needed to be succinct so that respondents could read them and answer questions within the 75 minute time frame. This meant the degree of fidelity to, in particular, the V\&B style, which takes its creators an entire book to explain, was clearly missing the richness of industry-standard architecture documents. While a conventional architecture document in V\&B style might be tens or hundreds of pages long, this is clearly infeasible as an experiment. 
We feel the documents created capture the structured nature of V\&B documents without overwhelming the reader. Arguably by simplifying the V\&B approach we made it easier to read and answer questions.

In addition, there are many uses for architecture descriptions, including prescribing what to build, analyzing system qualities and non-functional requirements, and educating stakeholders \citep{Clements:2010aa}. We focused the documents we created on another use, namely \textit{communicating design information}. We argue this is more relevant for developers and managers responsible for onboarding newcomers to a software project.

\subsubsection*{Internal Validity} The main sources of potential confounding factors originate from our pool of participants. One of the main strengths of our study was our ability to experiment with participants with two clearly distinct levels of familiarity with the underlying system. The trade-off is that the difference in student background also acts as a potential source of interference. Students at {\A} had used the reference system {\tool} during their earlier coursework, and were more familiar with the underlying purpose of the system and its goals. Students at {\B} did not get exposed to {\tool}, but were taught to document systems using the Views and Beyond style. While we investigated the possible effects of different classroom context on the data, there may be additional, unknown, effects besides the ones we controlled for. Nevertheless, we conducted the study at both institutions in exactly the same conditions to avoid threats related to variations in the experimental procedure. 

Another potential threat is the \textit{observer-expectancy effect}. Students were aware of the document style they had, and that it was being compared to a different style. It is possible that they may have tried to infer what style the instructor might prefer. We estimate this risk to be minimal for three reasons: \textit{a)} we collected the data anonymously, which offers the greatest level of privacy protection for such studies; \textit{b)} participants had to explicitly opt-in for this data to be utilized; and \textit{c}) our participants were senior students, not usually prone to be intimidated by in-class exercises.

We had a non-participation rate of 18\%. 
Thus it is possible that some of the rationale for non-participation was due to factors such as English language ability or lack of experience. However, since all students were required to participate in the exercise (but could refuse consent for data collection), we doubt these are the major factors. In particular consent was collected at arms-length to prevent power-over problems.

Our Bayesian analysis compared four models combining our independent variables, but other combinations of the variables are possible. We judged these models to best capture the exploratory concepts of interest. We provide a replication package for others to experiment with other model combinations.

\subsubsection*{External Validity} Our observations must be interpreted in the context in which they occur, namely a mixed senior-year undergraduate and graduate student class. We provide a detailed summary of the characteristics of our participants in Table~\ref{tbl:background2} to clarify this context. Although the particular controlled setting of our study required us to recruit our participants as students enrolled in a course, we note that 74\% of the participants had worked as interns or professional developers. In that respect their background characteristics is expected to be similar to, or overlap with, those of the target population of junior developers.

Because we limited the study to a single class session, we were not able to give participants what might be considered a ``realistic'' architecture document, for example with tens of pages. Similarly, the tasks were designed to be completed in a 10-15 minute time frame. We cannot claim that our results would apply directly to onboarding approaches that are often days-long camps, examining longer software architecture documents with more complex contexts and tasks. However, our use of a sequence of questions with different formats and progressive difficulty means our observations can be interpreted at a level of granularity that enables reasoning about the applicability of our results. Finally, with 65 participants and a response rate of 82\% across two institutions, we are confident that the data we collected is from participants that are valid representatives of senior students in computer science and software engineering.

\section{Conclusion}
\label{s:conclusion}

Software architecture documents serve many purposes. One of those purposes is to disseminate knowledge to newcomers to a software project. Our study explored to what extent the \textit{format} of architectural documentation matters when onboarding junior newcomers onto a project. In particular, we considered the impact of documentation format in relation to factors that include prior familiarity with the system and nature of the information needs fulfilled by the architecture documentation. 

By using a regression model to relate the quality of answers to question about the architecture of a system to other factors, we observed that, in our study context, the format of architectural documentation appears to be inconsequential for short architecture understanding tasks on a comparatively small system using a modest-sized document. The experimental protocol precluded an investigation with an extensive architectural description, for example in the hundreds of pages. It is possible that the effect of documentation format may be amplified by the length of the architecture description document. Nevertheless, our study setting remains relevant, as numerous software projects have a limited amount of architectural documentation. The implication of our observation is that, for on-boarding newcomers into a software project, documentation format is a second-order concern trumped in importance by giving newcomers a modicum of practical experience with the code.

We corroborated this general observation with two additional analyses that used different types of data from the study. In a second analysis, we examined to what extent participants relied on an investigation of source code to complement their reading of software architecture documentation. There again, we observed no significant difference between the documentation formats. Instead, we noticed that use of source code is strongly associated with the requirement to answer more open-ended questions with a focus on design.

In a third analysis, we turned to the subjective opinions of the participants about the documentation, as collected using Likert-scale questions. In this case again we found no clear preference for one format over the other.

We point out that the fact that we did not reliably observe that the format of architectural documentation mattered in our study does not mean that documentation format has no impact. As usual in experimental work, this impact may exist without having been detectable by our study design. In particular, such an impact may only manifest itself beyond a certain scale of  system or architectural documentation. 
Future work may be able to observe an effect of documentation format, although it is not clear what kind of study would accommodate the interactions of research participants with a huge system and massive architectural documents. 

In addition to contradicting the hypothesis that the format of architectural documentation matters (in our context), our study surfaced two factors that \textit{do} seem to matter: basic familiarity with the source code, and the type of information sought from the architectural documentation.  

\section*{Data Availability Statement}
Replication data for this paper, including the experimental artifacts and analysis scripts, are available at Zenodo, as part of this record: \replPackage{}.

\section*{Acknowledgements}
We thank Omar Elazhary for his feedback on the study and assistance with the exercise, and Jin Guo and Eirini Kalliamvakou for help with data collection. We are also grateful to Mathieu Nassif, Deeksha Arya, and the anonymous reviewers for their constructive feedback. This work was funded by NSERC.

\bibliographystyle{spbasic} 
\bibliography{references}

\end{document}